\documentclass[a4paper,12pt]{article}
\usepackage[pctex32]{graphics}
\usepackage{amssymb,amsmath}
\usepackage{amsmath,amssymb}
\usepackage{latexsym}
\usepackage{epsfig}
\usepackage[english]{babel}

\newcommand{\be}{\begin{equation}}
\newcommand{\ee}{\end{equation}}
\newcommand{\ba}{\begin{eqnarray}}
\newcommand{\ea}{\end{eqnarray}}

\begin{document}

\begin{titlepage}

\vspace{5mm}

\begin{center}

{\Large \bf Instability of  a Kerr  black hole \\ in f(R) gravity}

\vskip .6cm

\centerline{\large
 Yun Soo Myung$^{a}$}

\vskip .6cm

{Institute of Basic Science and Department of Computer Simulation,
\\Inje University, Gimhae 621-749, Korea \\}

\end{center}

\begin{center}
\underline{Abstract}
\end{center}

We study the stability of a rotating (Kerr) black hole in the viable
$f(R)$ gravity. The linearized-Ricci scalar equation shows the
superradiant instability, leading to the instability of the Kerr
black hole in $f(R)$ gravity.\vskip .6cm


\vskip 0.8cm

\vspace{15pt} \baselineskip=18pt

\noindent $^a$ysmyung@inje.ac.kr \\

\thispagestyle{empty}
\end{titlepage}

\newpage
%%%%%%%%%%%%%%%%%%%%%%%%%%%%%%%%%%%%%%%%%%%%%%%%%%%%%%%%%%%%%%%%%%
\section{Introduction}

 One of modified gravity theories, $f(R)$
gravity~\cite{NO,sf,FT,Nojiri:2010wj} has  much attentions as a
strong candidate for explaining the current and future accelerating
phases in the evolution of  universe~\cite{SN1,SN2}.   On the other
hand, the Schwarzschild-de Sitter black hole was firstly obtained
for a constant curvature scalar from $f(R)$ gravity~\cite{CENOZ}. A
Schwarzschild-anti de Sitter  black hole solution was obtained from
$f(R)$ gravity by requiring a negatively  constant curvature
scalar~\cite{delaCruzDombriz:2009et}. The trace of stress-energy
tensor should be   zero  to obtain a constant curvature black hole
when $f(R)$ gravity couples with matter of the Maxwell field
~\cite{delaCruzDombriz:2009et}, the Yang-Mills
field~\cite{Moon:2011hq}, and a nonlinear Maxwell
field~\cite{Sheykhi:2012zz}.

Most of  astrophysical black holes including supermassive black
holes are considered  to be a rotating black hole.  A rotating black
hole solution~\cite{Kerrsol} should be stable against the external
perturbations because it stands as a realistic object in the
sky~\cite{KZ}. The stability analysis of  the Kerr black hole is not
as straightforward as one has  performed the stability analysis of a
spherically symmetric Schwarzschild black
hole~\cite{Regge:1957td,Zerilli:1970se,Vishveshwara:1970cc} because
it is an axis-symmetric black hole.  Here  we would like to mention
that the stability analysis is based on the linearized equations and
thus, it does not guarantee the stability of black holes at the
nonlinear level. The Kerr black hole has been proven to be stable
against a massless graviton~\cite{PT,TP,Whit} and a massless
scalar~\cite{DI}. However, there exist the superradiant instability
(the black-hole bomb) when one chooses   a massive
scalar~\cite{ZE,CDLY,HH,Dol,Witek:2012tr,Dolan:2012yt} and a massive
vector~\cite{Pani:2012vp}. For  example of $f(R)=R+h R^2$, the Kerr
black hole is unstable because it could be transformed into a
massive scalar-tensor theory~\cite{Hersh:1985hz}.

It is known that the Kerr solution could be obtained from a limited
form (\ref{fform}) of  $f(R)$ gravity~\cite{kerr}. Interestingly, it
was shown  that a perturbed Kerr black hole could distinguish
Einstein gravity from  $f(R)$ gravity~\cite{BS}.  However, the
stability analysis of $f(R)$-rotating black hole is a formidable
task because $f(R)$ gravity contains  fourth-order derivative terms
in the linearized equation.   Transforming  the limited form of
$f(R)$ gravity into the scalar-tensor theory might resolve
difficulty, which leads   to the fact that the $f(R)$-rotating black
hole is unstable against a massive scalar perturbation when  one
used the black-hole bomb idea~\cite{Myung:2011we}.

In this work, we examine  the stability of a rotating  black hole in
the viable   $f(R)$ gravity.  We consider  the linearized Ricci
scalar  as a truly  massive spin-0 graviton  propagating on the Kerr
black hole spacetimes. Solving  its linearized equation shows a
superradiant instability, which dictates  the instability of the
Kerr black hole in $f(R)$ gravity.  This will be compared to the
Gregory-Laflamme instability of the massive spin-2 graviton in the
dRGT massive gravity~\cite{Babichev:2013una,Brito:2013wya} and the
fourth-order gravity~\cite{Myung:2013doa}.

\section{$f(R)$-rotating black holes}
We start with  the $f(R)$ gravity action
\begin{eqnarray}
S_{f}=\frac{1}{2\kappa^2}\int d^4 x\sqrt{-g} f(R)\label{Action}
\end{eqnarray}
with $\kappa^2=8\pi G$.  The Einstein equation  takes the form
\begin{eqnarray} \label{equa1}
R_{\mu\nu} f'(R)-\frac{1}{2}g_{\mu\nu}f(R)+
\Big(g_{\mu\nu}\nabla^2-\nabla_{\mu}\nabla_{\nu}\Big)f'(R)=0,
\end{eqnarray}
where the prime (${}^{\prime}$) denotes the differentiation with
respect to its argument.  It is well-known that Eq. (\ref{equa1})
provides  a solution with constant curvature scalar $R=\bar{R}$.  In
this case, Eq. (\ref{equa1}) reduces to
\begin{eqnarray} \label{equ1}
\bar{R}_{\mu\nu} f'(\bar{R})-\frac{1}{2}\bar{g}_{\mu\nu}f(\bar{R})=0
\end{eqnarray}
and thus,  the trace of (\ref{equ1})  determines the constant
curvature scalar to be
\begin{eqnarray}
\bar{R}=\frac{2f(\bar{R})}{f'(\bar{R})}\equiv 4\Lambda_f
\label{eqCR}
\end{eqnarray}
with $\Lambda_f$ the cosmological constant. The subscript `$f$'
denotes that the $\Lambda_f$ arose  from the $f(R)$ gravity.
Substituting this expression into (\ref{equ1}) leads to the Ricci
tensor
\begin{equation}
\bar{R}_{\mu\nu}=\frac{f(\bar{R})}{2f'(\bar{R})}\bar{g}_{\mu\nu}=\Lambda_f
\bar{g}_{\mu\nu}.
\end{equation}

To find the Kerr black hole solution with $\Lambda_f=0$
($\bar{R}_{\mu\nu}=\bar{R}=0$), one requires $f(0)=0$ with
$f'(0)\not=0$.  To this end, one has  to choose a specific form of
$f(R)$ as~\cite{kerr}
\begin{equation}
\label{fform} f(R)=a_1R+a_2R^2+a_3R^3+\cdots.
\end{equation}
In  Table 1, we list viable models of $f(R)$ gravity  which provide
the form (\ref{fform}).  Hence,  a model of $f(R)=R-\mu^4/R$ could
not provide a rotating black
hole~\cite{Carroll:2003wy,Dolgov:2003px} because $f(0)\to -\infty$
and $f'(0) \to \infty$. Also, the form of
$f(R)=\alpha\sqrt{R+\beta}$~\cite{Mazharimousavi:2012cb,Mazharimousavi:2012ca}
is  excluded because $f(0)=\alpha\sqrt{\beta}\not=0$. By the same
token, the two models of $f(R)=R^pe^{q/R}$ and $f(R)=R^p(\ln [\alpha
R])^q$~\cite{Leon:2013bra} are not suitable  for seeking the Kerr
black hole solution.

\begin{center}
\begin{table}
\begin{tabular}{|c|c|c|c|c|}
  \hline
   viable $f(R)$ gravity & $f(0)$ & $f'(0)$ & $f''(0)$ & $m^2_f=\frac{f'(0)}{3f''(0)}$  \\
  \hline
  $f_{\rm Q}=R+h R^2$\cite{Hersh:1985hz} & 0 & 1 & $2h$ & $ \frac{1}{6h}$\\ \hline
  $f_{\rm pE}^{p=1}=R^p e^{qR}$\cite{Leon:2013bra} & 0 & 1 & $2q$ & $\frac{1}{6q}$ \\ \hline
  $f_{\rm S}=R+\lambda R_s[(1+R^2/R_s^2)^{-n}-1]$\cite{Starobinsky:2007hu} & 0 & 1 & $-\frac{2 n\lambda }{R_s}$ & $-\frac{ R_s}{6 n \lambda }$\\ \hline
  $f^{n=1}_{\rm
  HS}=R-m^2\frac{c_1(R/m^2)^n}{1+c_2(R/m^2)^n}$\cite{Hu:2007nk}&0&$1-c_1$&$\frac{2c_1c_2}{m^2}$&
  $\frac{m^2}{6c_1c_2}$
  \\
  \hline
  $f^{n=2}_{\rm
  HS}=R-m^2\frac{c_1(R/m^2)^n}{1+c_2(R/m^2)^n}$\cite{Hu:2007nk}&0&1&$-\frac{2c_1}{m^2}$&$-\frac{m^2}{6c_1}$
  \\ \hline
  $f^{n>2}_{\rm
  HS}=R-m^2\frac{c_1(R/m^2)^n}{1+c_2(R/m^2)^n}$\cite{Hu:2007nk}&0&1&0& N/A
  \\ \hline
  $f_{\rm
  nE}=R-2\Lambda(1-e^{-R/\Lambda})$\cite{Cognola:2007zu}&0&1&$\frac{2}{\Lambda}$&$\frac{\Lambda}{6}$
  \\ \hline
  \end{tabular}
  \caption{Viable models of $f(R)$ gravity to provide the Kerr black hole as a solution. The condition of $m^2_f >0$ might be  different from that of a viable $f(R)$ gravity
  to explain the accelerating universe~\cite{FT}.}
\end{table}
\end{center}
 In this work,
we use the Boyer-Lindquist coordinates to represent an
axis-symmetric Kerr black hole solution with mass $M$ and angular
momentum $J$~\cite{Kerrsol}
\begin{eqnarray}
ds^2_{\rm Kerr}&=&\bar{g}_{\mu\nu}dx^\mu dx^\nu= -\left (
1-\frac{2Mr}{\rho^2}\right )dt^2
-\frac{2Mr a\sin^2\theta}{\rho^2}\, 2 dt d\phi  \nonumber \\
& & +\frac{\rho^2}{\Delta}\,dr^2 +\rho^2 d\theta^2+\left (
r^2+a^2+\frac{2Mr
a^2\sin^2\theta}{\rho^2}\right )\sin^2\theta\, d\phi^2 \, \nonumber \\
& &
 \label{Kerr}
\end{eqnarray}
with
\begin{eqnarray}
\Delta=r^2+a^2-2Mr,~ \rho^2=r^2+a^2 \cos^2\theta,~ a=\frac{J}{M}.
 \label{metric parameters}
\end{eqnarray}
Here we  use Planck units of $G=c=\hbar=1$ and thus, the mass $M$
has a length scale.  In the nonrotating limit of $a\to 0$,
(\ref{Kerr}) recovers the Schwarzschild black hole, while the
 limit of $a\to1$ corresponds  to the extremal Kerr black
hole. From  the condition of $\Delta=0(g^{rr}=0)$, we determine two
horizons  which are located at
\begin{equation}
r_\pm=M\pm \sqrt{M^2-a^2}.
\end{equation}
The angular velocity at the event horizon takes the form
\begin{equation}
\Omega=\frac{a}{2 M r_+}=\frac{a}{r_+^2+a^2}. \label{hav}
\end{equation}

In general,  one introduces the metric perturbation around the Kerr
black hole to study the stability of the black hole
\begin{eqnarray} \label{m-p}
g_{\mu\nu}=\bar{g}_{\mu\nu}+h_{\mu\nu}.
\end{eqnarray}
Hereafter we denote the background quantities with the ``overbar''
($\bar{R}_{\mu\nu}=0,~\bar{R}=0$). The Taylor expansions around the
zero  curvature scalar background is employed  to define the
linearized Ricci scalar~\cite{Myung:2011ih} as
\begin{eqnarray}
f(R)&=& f(\bar{R})+f'(\bar{R})\delta R(h)+\cdots, \\
f'(R)&=& f'(\bar{R})+f''(\bar{R})\delta R(h)+\cdots.
\end{eqnarray}
The linearized equation to (\ref{equa1}) is given by
\begin{eqnarray}
\delta R_{\mu\nu}(h)+\frac{f''(0)}{f'(0)}
\Bigg[-\frac{f'(0)}{2f''(0)}\bar{g}_{\mu\nu}+\bar{g}_{\mu\nu}\bar{\nabla}^2-\bar{\nabla}_{\mu}\bar{\nabla}_{\nu}
\Bigg]\delta R(h)=0,\label{leq}
\end{eqnarray}
where the linearized Ricci tensor and  scalar could be expressed in
terms of $h_{\mu\nu}$ as
\begin{eqnarray}
&&\delta
R_{\mu\nu}(h)=\frac{1}{2}\Bigg[\bar{\nabla}^{\rho}\bar{\nabla}_{\mu}h_{\nu\rho}+
\bar{\nabla}^{\rho}\bar{\nabla}_{\nu}h_{\mu\rho}-\bar{\nabla}^2h_{\mu\nu}-\bar{\nabla}_{\mu}
\bar{\nabla}_{\nu}h\Bigg],\label{lRmunu}\\
&&\delta R(h)=\bar{\nabla}^{\rho}\bar{\nabla}^{\sigma}h_{\rho\sigma}
-\bar{\nabla}^{2}h.\label{lR}
\end{eqnarray}
Considering (\ref{lRmunu}) and (\ref{lR}), the linearized equation
(\ref{leq}) is a fourth-order differential equation  with respect to
the metric perturbation $h_{\mu\nu}$, which is not a tractable
equation to be solved.  Choosing the Lorentz gauge of
$\bar{\nabla}_\nu h^{\mu\nu}=\bar{\nabla}^\mu h/2$ and using the
trace-reversed perturbation of $\tilde{h}_{\mu\nu}=h_{\mu\nu}-h
\bar{g}_{\mu\nu}/2$, equation (\ref{leq}) takes a relatively simple
from~\cite{BS}
\begin{eqnarray} \label{speq}
\bar{\nabla}^2\tilde{h}_{\mu\nu}+2\bar{R}_{\mu\rho\nu\sigma}\tilde{h}^{\rho\sigma}
+\frac{1}{3m^2_f}\Big(\bar{g}_{\mu\nu}\bar{\nabla}^2-\bar{\nabla}_{\mu}
\bar{\nabla}_{\nu}\Big)\bar{\nabla}^2\tilde{h}=0,
\end{eqnarray}
where the mass squared $m^2_f$  is defined by
\begin{equation}
m^2_f=\frac{f'(0)}{3f''(0)} >0.
\end{equation}
In case of Einstein gravity [$f(R)=R,f'(0)=1,f''(0)=0$],  Eq.
(\ref{speq}) leads to a well-known second-order equation for
$\tilde{h}_{\mu\nu}$ since the last fourth-order term is decoupled
from (\ref{speq}). However, we could not solve (\ref{speq}) directly
for  $m^2_f\not=\infty$ because it is a coupled fourth-order
equation for $\tilde{h}_{\mu\nu}$ and $\tilde{h}$.

\section{Superradiant instability of  Ricci scalar}
 It
is well-known that $f(R)$ gravity has 3 degrees of freedom (DOF)
without ghost in Minkowski spacetimes: 2 DOF for a massless spin-2
graviton and 1 DOF for a massive spin-0 graviton. The massive spin-0
graviton is usually described by the trace $h$ of $h_{\mu\nu}$, but
it could be represented by the linearized Ricci scalar  $\delta R$
because  $\delta R=-\bar{\nabla}^2h/2=\bar{\nabla}^2\tilde{h}/2$
under the Lorentz gauge.

 For this purpose,
 we may take
the trace of $(\ref{leq})$ with $\bar{g}^{\mu\nu}$. Then, we have a
massive equation for $\delta R$
\begin{eqnarray}
\Big(\bar{\nabla}^2-m^2_f\Big)\delta R=0\label{leq2}
\end{eqnarray}
which is considered as a second-order equation that describes the
linearized Ricci scalar  propagating on the background of Kerr black
hole.  In the previous work~\cite{Myung:2011we}, we have replaced
$\delta R$ by a scalaron $\delta A$ which could be interpreted  to
be a massive  scalar  in the scalar-tensor theory. This result is
meaningful only  if the scalaron approach (the scalar-tensor
 theory) represents $f(R)$ gravity truly.   However, it is noted that the linearized Ricci scalar by itself is regarded as a
physically propagating scalar because the $f(R)$ gravity includes  a
massive scalar graviton with single DOF.   In Table 1, we list
$m^2_f$ for viable $f(R)$ models. In order to not have a tachyonic
scalar, it should be positive ($m^2_f>0$) which implies that
$f'(0)>0$ and $f''(0)>0$.   Thus, one requires either $\lambda<0$ or
$n<0$ for the Starobinsky model ($f_{\rm
S}$)~\cite{Starobinsky:2007hu}. Also, $0<c_1<1$ is required for the
$n=1$ Hu-Sawiciki model ($ f^{n=1}_{\rm HS}$) and $c_1<0$ for the
$n=2$ Hu-Sawiciki model ($ f^{n=2}_{\rm HS}$)~\cite{Hu:2007nk}.
However, these are not mandatory  to explain the accelerating
universe when one uses  viable $f(R)$ gravity~\cite{FT}.

Reminding  the axis-symmetric background (\ref{Kerr}), it is
convenient to separate the linearized Ricci scalar into~\cite{Teu}
\begin{equation}
\delta R(t,r,\theta,\phi)=e^{-i\omega t + i m \phi} S^m
_l(\theta)u(r)\,, \label{sep}
\end{equation}
where $S^m _l(\theta)$ are spheroidal angular functions  with $l$
the spheroidal harmonic index  and  $m$ the azimuthal harmonic
number. Also, we choose a positive frequency $\omega$ of the mode
here. Plugging (\ref{sep}) into the linearized massive equation
(\ref{leq2}), one has  the angular and radial equations for $S^m
_l(\theta)$ and $u(r)$ as
\begin{eqnarray}
&& \frac{1}{\sin \theta}\partial_{\theta}\left ( \sin \theta
\partial_{\theta} S^m _l \right )  \nonumber \\
&&\hspace{1cm}+ \left [  a^2 (\omega^2-m^2_f) \cos^2
\theta-\frac{m^2}{\sin ^2{\theta}}+A_{lm} \right ]S^m _l =0\,,
\label{wave eq separated1}
\\
&& \Delta\partial_r \left ( \Delta \partial_r u \right )+ \Big[
\omega^2(r^2+a^2)^2-4M a m
\omega r +a^2 m^2 \nonumber \\
&& \hspace{3.2cm} - \Delta (a^2\omega^2+ m^2_f r^2+A_{lm}) \Big]
u=0\,,
 \label{wave eq separated}
\end{eqnarray}
where $A_{lm}$ is the separation constant whose form is given
by~\cite{BPT,HH}
\begin{eqnarray}
A_{lm}=l(l+1)+\sum^\infty_{k=1}c_ka^{2k}(m^2_f-\omega^2)^k
 \label{eigenvalues}
\end{eqnarray}
for $\omega \simeq m_f$.

 The radial Teukolsky equation takes the
Schr\"odinger form
\begin{equation}
-\frac{d^2\psi}{dy^2}+V(r,\omega)\psi=\omega^2\psi,~~\psi(r)=\sqrt{r^2+a^2}~u,
\end{equation}
where the tortoise  coordinate $y$ is defined by $dy=
\frac{r^2+a^2}{\Delta}dr$ and a $\omega$-dependent potential
$V_\omega(r)$ is given by
\begin{eqnarray}
V_\omega(r)&=&\frac{\Delta
m^2_f}{r^2+a^2}+\frac{4Mram\omega-a^2m^2+\Delta[A_{lm}+(\omega^2-m^2_f)a^2]}{(r^2+a^2)^2}
 \nonumber \\
 &+&\frac{\Delta(3r^2-4Mr+a^2)}{(r^2+a^2)^3}-\frac{3\Delta^2r^2}{(r^2+a^2)^4}.
\end{eqnarray}
Its asymptotic form is given by
\begin{equation}
V_\omega \to \omega^2-m^2_f,~~y\to \infty~(r\to\infty),
\end{equation}
and its form near the event horizon is
\begin{equation}
V_\omega \to (\omega-m\Omega)^2,~~y\to -\infty~(r\to r_+).
\end{equation}
Here we impose  the two boundary conditions of purely ingoing waves
near the horizon and a decaying (bounded) solution at spatial
infinity. These are known to be boundary conditions for quasibound
states~\cite{ZE}. Near the horizon and at the spatial infinity, the
linearized Ricci scalar takes the form
\begin{eqnarray}
\psi &\sim & e^{- i(\omega-m\Omega)
y}\,\,,\,\,y \to -\infty \label{bc1}\\
\psi &=& e^{-\sqrt{m^2_f-\omega^2}y},~~y\rightarrow \infty.
\label{bc2}
\end{eqnarray}
 Then, we may choose  an ingoing mode near the horizon
\begin{equation}
[e^{-i\omega t}\psi]_{\rm in}\sim e^{-i\omega t}e^{-
i(\omega-m\Omega) y}. \nonumber \end{equation} From (\ref{bc2}), a
bound state of  exponentially decaying mode at spatial infinity  is
characterized by the condition
\begin{equation} \label{bc3}
\omega^2<m^2_f.
\end{equation}
The three boundary conditions (\ref{bc1})-(\ref{bc3}) imply  a
discrete set of resonances $\{\omega_n\}$ which corresponds to bound
states of the linearized Ricci scalar.

In addition, let us consider  a wave of  $e^{-i\omega t}e^{im\phi}$
with $m>0$ and real $\omega$  which is propagating into a rotating
black hole with angular velocity $\Omega$. If the frequency of the
incident wave satisfies the condition~\cite{PT}
\begin{equation} \label{sr-con}
\omega <m\Omega,
\end{equation}
then the scattered wave is amplified. This is called  the
superradiance condition for a bosonic field~\cite{BPT}.

The existence of superradiant modes can be converted into an
instability of the black hole background if a mechanism to trap
these modes in a vicinity of the black hole  is provided.  There are
two mechanisms to achieve it.   If one surrounds the black hole by
putting a reflecting mirror, the wave will bounce back and forth
between black hole and mirror, amplifying itself each time and
eventually producing a nonnegligible backreaction on the black hole
background. This yields  an exponentially growing mode which can be
no longer considered as a perturbation, demonstrating the
instability of the black hole. Secondly, the nature may provide its
own mirror when one introduces a massive scalar. Press and Teukolsky
have suggested to use this mechanism to define the black-hole
bomb~\cite{PT} by introducing  a massive scalar with mass ${\cal M}$
propagating around the Kerr black hole with mass $M$. For
$\omega<{\cal M}(\omega^2<{\cal M}^2)$, the mass term works as a
mirror effectively. The maximum growth rate for the instability is
associated with modes with $\omega=\omega_R+i\omega_I$. The sign of
$\omega_I$ usually determines whether the solution is decaying
($\omega_I<0$) or growing ($\omega_I>0$) in the time evolution. It
was shown that $\omega_I M \sim 6 \times 10^{-5}$ for mirrorlike
boundary conditions~\cite{CDLY} and $\omega_I M \sim 1.72 \times
10^{-7}$ for massive scalars~\cite{Dolan:2012yt}. Here the growth
time scale is given by $\tau=1/\omega_I$.

More explicitly, according to the Hod's argument~\cite{Hod:2012zza},
two ingredients are necessary  to trigger the instability of the
Kerr black hole when one uses a massive scalar perturbation: 1) The
existence of an ergoregion where superradiant amplification of the
waves takes place.  2) The existence of a trapping potential well
($\sim$) for quasibound states is between the potential barrier from
ergoregion and potential barrier from the mass (see Fig. 15 of
Ref.\cite{KZ} and Fig. 7 of Ref.\cite{Arvanitaki:2010sy}). The first
ingredient is usually implemented  by the superradiance condition
(\ref{sr-con}). The second ingredient is supplied   by  the
condition of the bound states for modes in the regime
\begin{equation} \label{se-i}
\frac{m^2_f}{2}<\omega^2< m^2_f.
\end{equation}
Combining (\ref{sr-con}) with (\ref{se-i}), one finds a restricted
regime for the mass
\begin{equation}
m_f <\sqrt{2} \omega <\sqrt{2} m \Omega
\end{equation}
which implies an  inequality between mass $m_f$ of the Ricci scalar
and the angular velocity $\Omega$ of the rotating black hole
\begin{equation} \label{ineq}
m_f <\sqrt{2} m \Omega
\end{equation}
which is the main result of our work.

 The bound (\ref{ineq})  is
reminiscent of the Gregory-Laflamme $s$-mode instability
~\cite{Gregory:1993vy} for a massive spin-2 graviton  with mass
$\mu$ propagating on the spherically symmetric Schwarzschild black
hole spacetimes (mass $2M_{\rm S}=r_0$).  Choosing the
transverse-traceless gauge of $\bar{\nabla}^\mu h_{\mu\nu}=0$ and
$h=0$, its linearized equation takes the form
\begin{equation}
\bar{\nabla}^2
h_{\mu\nu}+2\bar{R}_{\alpha\mu\beta\nu}h^{\alpha\beta}-\mu^2h_{\mu\nu}=0.
\end{equation}
which describes  5 DOF of a massive spin-2 propagating on the
Schwarzschild black hole spacetimes.
 To
this end, we would like to mention that the stability of the
Schwarzschild  black hole in four-dimensional massive gravity is
determined by using the Gregory-Laflamme instability of a
five-dimensional black string. It turned out that the small
Schwarzschild black holes in the dRGT massive
gravity~\cite{Babichev:2013una,Brito:2013wya} and fourth-order
gravity~\cite{Myung:2013doa} are  unstable against the metric and
Ricci tensor perturbations because the inequality is satisfied as
\begin{equation}
\mu \le \frac{0.438}{M_{\rm S}}.
\end{equation}

On the other hand, the dynamics of Ricci  scalar with mass $m_f$ is
expected to be stable in the complementary regime
\begin{equation}
m_f \ge \sqrt{2} m\Omega.
\end{equation}
Similarly, the massive graviton is stable if  it propagates around
the large Schwarzschild black hole  which satisfies the
bound~\cite{Brito:2013xaa}
\begin{equation}
\mu >\frac{0.438}{M_{\rm S}}.
\end{equation}

\section{Discussions}

We  have  investigated   the stability of a rotating  black hole in
the viable $f(R)$ gravity explicitly.   Even though viable $f(R)$
gravity is promising to describe the current accelerating universe,
it does not have a room to accommodate a rotating black hole because
the Kerr black is unstable against the  Ricci scalar perturbation.
This superradiant instability (the black-hole bomb) arose from the
nature of $f(R)$ gravity which provides a massive scalar graviton
with single DOF,  in addition to a massless spin-2 graviton with 2
DOF. This implies strongly that the Kerr black holes do not exist in
$f(R)$ gravity and/or they do not form in the process of the $f(R)$
gravitational collapse~\cite{Babichev:2013una}.  On the other hand,
we expect from the scalar-tensor theory~\cite{Myung:2011ih} that the
Schwarzschild black hole is stable against the Ricci scalar
perturbation in viable $f(R)$ gravity because it is a nonrotating
black hole.

We summarize the type of black hole instabilities found in the dRGT
massive gravity, fourth-order gravity, and $f(R)$ gravity in Table
2. Let us compare  the instability condition of Kerr black hole in
$f(R)$ gravity with  the instability condition  of Schwarzschild
black hole in dRGT massive gravity and fourth-order gravity.  The
instability of the Schwarzschild  black hole in four-dimensional
massive gravity is determined by using the Gregory-Laflamme
instability of a five-dimensional black string. The two conditions
of $\mu \le \frac{0.438}{M_{\rm S}}$ and $m_2 \le \frac{1}{2M_{\rm
S}}$ imply that the small Schwarzschild black holes in the dRGT
massive gravity~\cite{Babichev:2013una,Brito:2013wya} and
fourth-order gravity~\cite{Myung:2013doa} are  unstable against the
$s$-mode metric and Ricci tensor perturbations. These instabilities
arose from the massiveness of $s$-mode spin-2 graviton propagating
on the non-rotating small black hole with mass $M_{\rm S}$. On the
other hand, the condition of $m_f <\sqrt{2}m \Omega$ arose from the
massiveness of spin-0 graviton with azimuthal number $m$ propagating
on the rotating black hole with angular velocity $\Omega$. Even
though the massiveness is a common factor for  both instabilities,
the phenomena of the instability are different: GL
 black string instability and black hole bomb.

Finally, we conclude   that  the massive graviton instabilities are
quite different from the Regge-Wheeler-Zerilli stability for a
massless graviton~\cite{Regge:1957td,Zerilli:1970se,TP}. It suggests
that a massive gravity is hard to possess  the black hole.
\begin{center}
\begin{table}
\begin{tabular}{|c|c|c|c|}
  \hline
  theory & MG~\cite{Babichev:2013una} & FOG~\cite{Myung:2013doa} & $f(R)$ gravity
  \\ \hline
  black hole & Schwarzschild  & Schwarzschild & Kerr  \\ \hline
  perturbation & metric tensor   & Ricci tensor  & Ricci scalar  \\
  & $h_{\mu\nu}$   & $\delta R_{\mu\nu}$ & $\delta R$ \\ \hline
  gauge-fixing & TT & TT& not imposed \\ \hline
  No.  of massive DOF & 5 & 5& 1 \\ \hline
  nature of instability  & GL & GL&  superradiance  \\ \hline
  condition of instability & $\mu \le \frac{0.438}{M_{\rm S}}$ & $m_2 \le \frac{1}{2M_{\rm S}}$ & $m_f <\sqrt{2}m \Omega$ \\
  \hline
\end{tabular}
\caption{ Type of black hole instabilities in the the dRGT massive
gravity (MG), fourth-order gravity (FOG),  and $f(R)$ gravity. TT
denotes the transverse-traceless gauge and GL represents the
Gregory-Laflamme black string. }
\end{table}
\end{center}

\section*{Acknowledgement}

This work was supported supported by the National Research
Foundation of Korea (NRF) grant funded by the Korea government
(MEST) (No.2012-R1A1A2A10040499).


\begin{thebibliography}{99}
\bibitem{NO}
  S.~Nojiri and S.~D.~Odintsov,
  %``Introduction to modified gravity and gravitational alternative for dark
  %energy,''
  eConf {\bf C0602061} (2006) 06
  [Int.\ J.\ Geom.\ Meth.\ Mod.\ Phys.\  {\bf 4}, 115 (2007)]
  [arXiv:hep-th/0601213].
  %%CITATION = 00436,4,115;%%


\bibitem{sf}
  T.~P.~Sotiriou and V.~Faraoni,
  %``f(R) Theories Of Gravity,''
  Rev.\ Mod.\ Phys.\  {\bf 82}, 451 (2010)
  [arXiv:0805.1726 [gr-qc]].
  %%CITATION = RMPHA,82,451;%%


\bibitem{FT}
  A.~De Felice and S.~Tsujikawa,
  %``f(R) theories,''
  Living Rev.\ Rel.\  {\bf 13}, 3 (2010)
  [arXiv:1002.4928 [gr-qc]].
  %%CITATION = 00222,13,3;%%

%\cite{Nojiri:2010wj}
\bibitem{Nojiri:2010wj}
  S.~'i.~Nojiri and S.~D.~Odintsov,
  %``Unified cosmic history in modified gravity: from F(R) theory to Lorentz non-invariant models,''
   Phys.\ Rept.\  {\bf 505}, 59 (2011)  [arXiv:1011.0544 [gr-qc]].
   %%CITATION = ARXIV:1011.0544;%%  %453 citations counted in INSPIRE as of 05 Sep 2013




  \bibitem{SN1}
  S.~Perlmutter {\it et al.}  [Supernova Cosmology Project Collaboration],
  %``Measurements of Omega and Lambda from 42 High-Redshift Supernovae,''
  Astrophys.\ J.\  {\bf 517} (1999) 565
  [arXiv:astro-ph/9812133].
  %%CITATION = ASJOA,517,565;%%

\bibitem{SN2}
 A.~G.~Riess {\it et al.}  [Supernova Search Team Collaboration],
  %``Observational Evidence from Supernovae for an Accelerating Universe and a
  %Cosmological Constant,''
  Astron.\ J.\  {\bf 116} (1998) 1009
  [arXiv:astro-ph/9805201].
  %%CITATION = ANJOA,116,1009;%%



\bibitem{CENOZ}
  G.~Cognola, E.~Elizalde, S.~Nojiri, S.~D.~Odintsov and S.~Zerbini,
  %``One-loop f(R) gravity in de Sitter universe,''
  JCAP {\bf 0502}, 010 (2005)
  [arXiv:hep-th/0501096].
  %%CITATION = JCAPA,0502,010;%%


  %\cite{delaCruzDombriz:2009et}
\bibitem{delaCruzDombriz:2009et}
  A.~de la Cruz-Dombriz, A.~Dobado and A.~L.~Maroto,
  %``Black Holes in f(R) theories,''
   Phys.\ Rev.\ D {\bf 80}, 124011 (2009)  [Erratum-ibid.\ D {\bf 83}, 029903 (2011)]  [arXiv:0907.3872 [gr-qc]].
    %%CITATION = ARXIV:0907.3872;%%  %49 citations counted in INSPIRE as of 05 Sep 2013


%\cite{Moon:2011hq}
\bibitem{Moon:2011hq}
  T.~Moon, Y.~S.~Myung and E.~J.~Son,
  %``f(R) black holes,''
  Gen.\ Rel.\ Grav.\  {\bf 43}, 3079 (2011)  [arXiv:1101.1153 [gr-qc]].
  %%CITATION = ARXIV:1101.1153;%%  %24 citations counted in INSPIRE as of 05 Sep 2013

%\cite{Sheykhi:2012zz}
\bibitem{Sheykhi:2012zz}
  A.~Sheykhi,
  %``Higher-dimensional charged $f(R)$ black holes,''
  Phys.\ Rev.\ D {\bf 86}, 024013 (2012)  [arXiv:1209.2960 [hep-th]].
  %%CITATION = ARXIV:1209.2960;%%  %6 citations counted in INSPIRE as of 05 Sep 2013

\bibitem{Kerrsol}
  R.~P.~Kerr,
  %``Gravitational field of a spinning mass as an example of algebraically
  %special metrics,''
  Phys.\ Rev.\ Lett.\  {\bf 11}, 237 (1963).
  %%CITATION = PRLTA,11,237;%%

%\cite{Konoplya:2011qq}
\bibitem{KZ}
  R.~A.~Konoplya and A.~Zhidenko,
  %``Quasinormal modes of black holes: From astrophysics to string theory,''
  Rev.\ Mod.\ Phys.\  {\bf 83}, 793 (2011)  [arXiv:1102.4014 [gr-qc]].
  %%CITATION = ARXIV:1102.4014;%%  %85 citations counted in INSPIRE as of 06 Sep 2013

%\cite{Regge:1957td}
\bibitem{Regge:1957td}
  T.~Regge and J.~A.~Wheeler,
  %``Stability of a Schwarzschild singularity,''
   Phys.\ Rev.\  {\bf 108}, 1063 (1957).
    %%CITATION = PHRVA,108,1063;%%
     %820 citations counted in INSPIRE as of 15 Mar 2013

%\cite{Zerilli:1970se}
\bibitem{Zerilli:1970se}
  F.~J.~Zerilli,
  %``Effective potential for even parity Regge-Wheeler gravitational perturbation equations,''
   Phys.\ Rev.\ Lett.\  {\bf 24}, 737 (1970).
     %%CITATION = PRLTA,24,737;%%
     %291 citations counted in INSPIRE as of 15 Mar 2013

  %\cite{Vishveshwara:1970cc}
\bibitem{Vishveshwara:1970cc}
  C.~V.~Vishveshwara,
  %``Stability of the schwarzschild metric,''
  Phys.\ Rev.\ D {\bf 1}, 2870 (1970).
   %%CITATION = PHRVA,D1,2870;%%
   %194 citations counted in INSPIRE as of 15 Mar 2013

\bibitem{PT}
  W.~H.~Press and S.~A.~Teukolsky,
  %``Floating Orbits, Superradiant Scattering and the Black-hole Bomb,''
  Nature {\bf 238}, 211 (1972).
  %%CITATION = NATUA,238,211;%%


  \bibitem{TP}
  S.~A.~Teukolsky and W.~H.~Press,
  %``Perturbations of a rotating black hole. III - Interaction of the hole with
  %gravitational and electromagnet ic radiation,''
  Astrophys.\ J.\  {\bf 193}, 443 (1974).
  %%CITATION = ASJOA,193,443;%

\bibitem{Whit}
  B.~F.~Whiting,
  %``MODE STABILITY OF THE KERR BLACK HOLE,''
  J.\ Math.\ Phys.\  {\bf 30}, 1301 (1989).
  %%CITATION = JMAPA,30,1301;%%

\bibitem{DI}
  S.~L.~Detweiler and J.~R.~Ipser,
  %``Stability of scalar perturbations of a Kerr-metric black hole,''
  Astrophys.\ J.\  {\bf 185}, 675 (1973).
  %%CITATION = ASJOA,185,675;%%


 \bibitem{ZE}
  T.~J.~M.~Zouros and D.~M.~Eardley,
  %``INSTABILITIES OF MASSIVE SCALAR PERTURBATIONS OF A ROTATING BLACK HOLE,''
  Annals Phys.\  {\bf 118}, 139 (1979).
  %%CITATION = APNYA,118,139;%%

\bibitem{CDLY}
  V.~Cardoso, O.~J.~C.~Dias, J.~P.~S.~Lemos and S.~Yoshida,
  %``The Black hole bomb and superradiant instabilities,''
  Phys.\ Rev.\  D {\bf 70}, 044039 (2004)
  [Erratum-ibid.\  D {\bf 70}, 049903 (2004)]
  [arXiv:hep-th/0404096].
  %%CITATION = PHRVA,D70,044039;

\bibitem{HH}
  S.~Hod and O.~Hod,
  %``Analytic treatment of the black-hole bomb,''
  Phys.\ Rev.\  D {\bf 81}, 061502 (2010)
  [arXiv:0910.0734 [gr-qc]].
  %%CITATION = PHRVA,D81,061502

  \bibitem{Dol}
  S.~R.~Dolan,
  %``Instability of the massive Klein-Gordon field on the Kerr spacetime,''
  Phys.\ Rev.\  D {\bf 76}, 084001 (2007)
  [arXiv:0705.2880 [gr-qc]].
  %%CITATION = PHRVA,D76,084001

  %\cite{Witek:2012tr}
\bibitem{Witek:2012tr}
  H.~Witek, V.~Cardoso, A.~Ishibashi and U.~Sperhake,
  %``Superradiant instabilities in astrophysical systems,''
  Phys.\ Rev.\ D {\bf 87}, 043513 (2013)  [arXiv:1212.0551 [gr-qc]].
   %%CITATION = ARXIV:1212.0551;%%  %16 citations counted in INSPIRE as of 06 Sep 2013

%\cite{Dolan:2012yt}
\bibitem{Dolan:2012yt}
  S.~R.~Dolan,
  %``Superradiant instabilities of rotating black holes in the time domain,''
   Phys.\ Rev.\ D {\bf 87}, 124026 (2013)  [arXiv:1212.1477 [gr-qc]].
    %%CITATION = ARXIV:1212.1477;%%  %13 citations counted in INSPIRE as of 06 Sep 2013

%\cite{Pani:2012vp}
\bibitem{Pani:2012vp}
  P.~Pani, V.~Cardoso, L.~Gualtieri, E.~Berti and A.~Ishibashi,
  %``Black hole bombs and photon mass bounds,''
  Phys.\ Rev.\ Lett.\  {\bf 109}, 131102 (2012)  [arXiv:1209.0465 [gr-qc]].
   %%CITATION = ARXIV:1209.0465;%%  %22 citations counted in INSPIRE as of 10 Sep 2013



%\cite{Hersh:1985hz}
\bibitem{Hersh:1985hz}
  J.~Hersh and R.~Ove,
  %``Instability Of The Kerr Solution Of Fourth Order Gravity,''
  Phys.\ Lett.\ B {\bf 156}, 305 (1985).
   %%CITATION = PHLTA,B156,305;%%  %13 citations counted in INSPIRE as of 06 Sep 2013


\bibitem{kerr}
  D. Psaltis, D. Perrodin, K. R. Dienes, and I. Mocioiu,
  %``Kerr Black Holes are Not Unique to General Relativity,''
  Phys.\ Rev.\ Lett. {\bf 100}, 091101 (2008)
  [arXiv:0710.4564 [astro-ph]].

  \bibitem{BS}
  E.~Barausse and T.~P.~Sotiriou,
  %``Perturbed Kerr Black Holes can probe deviations from General Relativity,''
  Phys.\ Rev.\ Lett.\  {\bf 101}, 099001 (2008)
  [arXiv:0803.3433 [gr-qc]].
  %%CITATION = PRLTA,101,099001;%%



  %\cite{Myung:2011we}
\bibitem{Myung:2011we}
  Y.~S.~Myung,
  %``Instability of rotating black hole in a limited form of $f(R)$ gravity,''
   Phys.\ Rev.\ D {\bf 84}, 024048 (2011)  [arXiv:1104.3180 [gr-qc]].
    %%CITATION = ARXIV:1104.3180;%%  %12 citations counted in INSPIRE as of 06 Sep 2013


%\cite{Babichev:2013una}
\bibitem{Babichev:2013una}
  E.~Babichev and A.~Fabbri,
  %``Instability of black holes in massive gravity,''
   Class.\ Quant.\ Grav.\  {\bf 30}, 152001 (2013)  [arXiv:1304.5992 [gr-qc]].
    %%CITATION = ARXIV:1304.5992;%%  %8 citations counted in INSPIRE as of 05 Aug 2013

%\cite{Brito:2013wya}
\bibitem{Brito:2013wya}
  R.~Brito, V.~Cardoso and P.~Pani,
  %``Massive spin-2 fields on black hole spacetimes: Instability of the Schwarzschild and Kerr solutions and bounds on the graviton mass,''
  Phys.\ Rev.\ D {\bf 88}, 023514 (2013)  [arXiv:1304.6725 [gr-qc]].
  %%CITATION = ARXIV:1304.6725;%%  %10 citations counted in INSPIRE as of 05 Sep 2013


%\cite{Myung:2013doa}
\bibitem{Myung:2013doa}
  Y.~S.~Myung,
  %``Stability of Schwarzschild black holes in fourth-order gravity revisited,''
 Phys.\  Rev.\ D {\bf 88}, 024039 (2013)  [arXiv:1306.3725 [gr-qc]].
 %%CITATION = ARXIV:1306.3725;%%



%\cite{Myung:2011ih}
\bibitem{Myung:2011ih}
  Y.~S.~Myung, T.~Moon and E.~J.~Son,
  %``Stability of f(R) black holes,''
  Phys.\ Rev.\ D {\bf 83}, 124009 (2011)  [arXiv:1103.0343 [gr-qc]].
   %%CITATION = ARXIV:1103.0343;%%  %14 citations counted in INSPIRE as of 06 Sep 2013

%\cite{Carroll:2003wy}
\bibitem{Carroll:2003wy}
  S.~M.~Carroll, V.~Duvvuri, M.~Trodden and M.~S.~Turner,
  %``Is cosmic speed - up due to new gravitational physics?,''
   Phys.\ Rev.\ D {\bf 70}, 043528 (2004)  [astro-ph/0306438].
    %%CITATION = ASTRO-PH/0306438;%%  %1159 citations counted in INSPIRE as of 06 Sep 2013



%\cite{Dolgov:2003px}
\bibitem{Dolgov:2003px}
  A.~D.~Dolgov and M.~Kawasaki,
  %``Can modified gravity explain accelerated cosmic expansion?,''
  Phys.\ Lett.\ B {\bf 573}, 1 (2003)  [astro-ph/0307285].
  %%CITATION = ASTRO-PH/0307285;%%  %394 citations counted in INSPIRE as of 06 Sep 2013


%\cite{Mazharimousavi:2012cb}
\bibitem{Mazharimousavi:2012cb}
  S.~H.~Mazharimousavi and M.~Halilsoy,
  %``Comment on 'Static and spherically symmetric black holes in f(R) theories',''
  Phys.\ Rev.\ D {\bf 86}, 088501 (2012)  [arXiv:1210.4699 [gr-qc]].
  %%CITATION = ARXIV:1210.4699;%%  %1 citations counted in INSPIRE as of 06 Sep 2013

%\cite{Mazharimousavi:2012ca}
\bibitem{Mazharimousavi:2012ca}
  S.~H.~Mazharimousavi, M.~Kerachian and M.~Halilsoy,
  %``Existence of Reissner-Nordstrom type black holes in f(R) gravity,''
   Int.\ J.\ Mod.\ Phys.\ D {\bf 22}, 1350057 (2013)  [arXiv:1210.4696 [gr-qc]].
    %%CITATION = ARXIV:1210.4696;%%  %2 citations counted in INSPIRE as of 06 Sep 2013


%\cite{Leon:2013bra}
\bibitem{Leon:2013bra}
  G.~Leon and A.~A.~Roque,
  %``Qualitative analysis of Kantowski-Sachs metric in a generic class of $f(R)$ models,''
  arXiv:1308.5921 [astro-ph.CO].  %%CITATION = ARXIV:1308.5921;%%


%\cite{Starobinsky:2007hu}
\bibitem{Starobinsky:2007hu}
  A.~A.~Starobinsky,
  %``Disappearing cosmological constant in f(R) gravity,''
  JETP Lett.\  {\bf 86}, 157 (2007)  [arXiv:0706.2041 [astro-ph]].
   %%CITATION = ARXIV:0706.2041;%%  %368 citations counted in INSPIRE as of 06 Sep 2013


%\cite{Hu:2007nk}
\bibitem{Hu:2007nk}
  W.~Hu and I.~Sawicki,
  %``Models of f(R) Cosmic Acceleration that Evade Solar-System Tests,''
  Phys.\ Rev.\ D {\bf 76}, 064004 (2007)  [arXiv:0705.1158 [astro-ph]].
  %%CITATION = ARXIV:0705.1158;%%  %552 citations counted in INSPIRE as of 06 Sep 2013



%\cite{Cognola:2007zu}
\bibitem{Cognola:2007zu}
  G.~Cognola, E.~Elizalde, S.~Nojiri, S.~D.~Odintsov, L.~Sebastiani and S.~Zerbini,
  %``A Class of viable modified f(R) gravities describing inflation and the onset of accelerated expansion,''
  Phys.\ Rev.\ D {\bf 77}, 046009 (2008)  [arXiv:0712.4017 [hep-th]].
  %%CITATION = ARXIV:0712.4017;%%  %220 citations counted in INSPIRE as of 06 Sep 2013




\bibitem{Teu}
  S.~A.~Teukolsky,
  %``Rotating black holes - separable wave equations for gravitational and
  %electromagnetic perturbations,''
  Phys.\ Rev.\ Lett.\  {\bf 29}, 1114 (1972).
  %%CITATION = PRLTA,29,1114;



\bibitem{BPT}
  J.~M.~Bardeen, W.~H.~Press and S.~A.~Teukolsky,
  %``Rotating black holes: Locally nonrotating frames, energy extraction, and
  %scalar synchrotron radiation,''
  Astrophys.\ J.\  {\bf 178}, 347 (1972).
  %%CITATION = ASJOA,178,347;%%


%\cite{Hod:2012zza}
\bibitem{Hod:2012zza}
  S.~Hod,
  %``On the instability regime of the rotating Kerr spacetime to massive scalar perturbations,''
   Phys.\ Lett.\ B {\bf 708}, 320 (2012)  [arXiv:1205.1872 [gr-qc]].
   %%CITATION = ARXIV:1205.1872;%%  %7 citations counted in INSPIRE as of 06 Sep 2013


%\cite{Arvanitaki:2010sy}
\bibitem{Arvanitaki:2010sy}
  A.~Arvanitaki and S.~Dubovsky,
  %``Exploring the String Axiverse with Precision Black Hole Physics,''
  Phys.\ Rev.\ D {\bf 83}, 044026 (2011)  [arXiv:1004.3558 [hep-th]].
  %%CITATION = ARXIV:1004.3558;%%  %47 citations counted in INSPIRE as of 06 Sep 2013


%\cite{Gregory:1993vy}
\bibitem{Gregory:1993vy}
  R.~Gregory and R.~Laflamme,
  %``Black strings and p-branes are unstable,''
   Phys.\ Rev.\ Lett.\  {\bf 70}, 2837 (1993)  [hep-th/9301052].
   %%CITATION = HEP-TH/9301052;%%  %593 citations counted in INSPIRE as of 14 Jun 2013


%\cite{Brito:2013xaa}
\bibitem{Brito:2013xaa}
  R.~Brito, V.~Cardoso and P.~Pani,
  %``Black holes with massive graviton hair,''
   Phys.\  Rev.\ D {\bf 88}, 064006 (2013)  [arXiv:1309.0818 [gr-qc]].
   %%CITATION = ARXIV:1309.0818;%%









 \end{thebibliography}
\end{document}